*Gedanken* experiment to test Bell's spaceship paradox

Len Zane, Emeritus Professor, University of Nevada, Las Vegas

Abstract

In Bell's spaceship paradox, a thread connects two spaceships moving with identical accelerations.  As the ships accelerate in lockstep, the tension in the string increases due to relativistic contraction and the string eventually breaks.  What happens if instead of exactly matched accelerations, the changing tension in the string introduces small fluctuations that disrupts the lockstep acceleration?  This relaxation of the requirement of lockstep acceleration dramatically changes the motion of the two ships.  At a surprisingly small velocity, the two ships begin to move in a manner that maintains a constant proper distance between them.  This allows the ships to accelerate indefinitely without breaking the string.

Bell's Paradox

The initial statement of what has become known as Bell's spaceship paradox was published in 1959[1].  The paradox became more widely known after Bell published his statement of the paradox in an article in 1987[2].  The following description of the paradox is from Bell's article:

*"Three small spaceships, A, B, and C, drift freely in a region of space remote from other matter, without rotation and without relative motion, with B and C equidistant from A*
*On reception of a signal from A the motors of B and C are ignited and they accelerate gently.*

*"Let ships B and C be identical, and have identical acceleration programmes. Then (as reckoned by an observer in A) they will have at every moment the same velocity, and so remain displaced one from the other by a fixed distance. Suppose that a fragile thread is tied initially between projections from B and C.  If it is just long enough to span the required distance initially, then as the rockets speed up, it will become too short, because of its need to Fitzgerald contract, and must finally break. It must break when, at a sufficiently high velocity, the artificial prevention of the natural contraction imposes intolerable stress."*

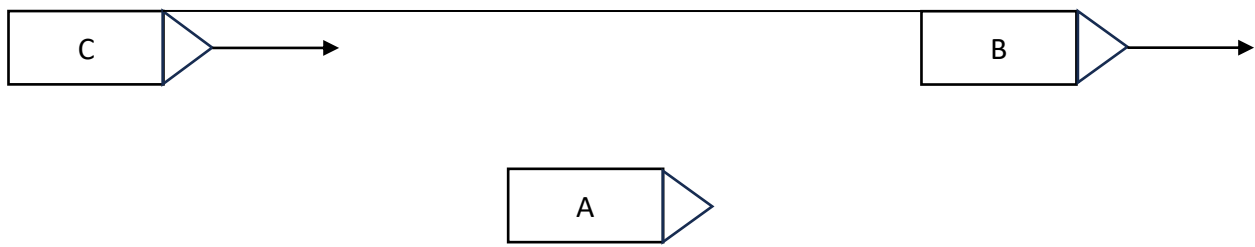

Figure 1:  Shows the orientation of the three ships when B and C begin to accelerate.

An easy way to understand the paradox is to imagine that the ships stop accelerating at an agreed upon velocity, v.  The coasting ships are now in an inertial reference frame moving with a constant velocity with respect to the initial inertial frame of spaceship A.  When viewed from the rest frame of ship A, the distance between ships B and C hasn't changed.  On the other hand, the distance measured in the frame where ships B and C are at rest and coasting with velocity v has increased by the factor $1/\sqrt{1 - v^2/c^2}$.  There is inevitably a stretching factor that will at some point cause the thread to break.

The *Gedanken* Experiment

The remainder of this article describes a *gedanken* experiment that shows that the paradox appears to be untestable assuming reasonable values for all the necessary parameters.  These reasonable values are listed in the next paragraph and were found by searching the internet.

The mass of the SpaceX starship is about 5 million kilograms.  This is used as the mass of spaceships B and C.  The model for the fragile thread is a three-ply cotton thread with a mass of 50 gm/1000 m.  The tensile strength is 2.2 pounds or 10 Newtons where g, the acceleration of gravity has been rounded up to 10 m/s².  That breaking tension is reached when the thread is stretched by 7%.  The ships are connected by a thread one kilometer long.  Ships B and C maintain a constant proper, or local, acceleration of 10 m/s².  Ship A remains stationary in the initial rest frame.

The experiment starts when the two spaceships begin their identical gentle accelerations in the same direction.  The proper acceleration of each ship slowly increases to 10 m/s² and then continues at that value.  Assume that the thread contracts only an infinitesimal amount during the time needed for the ships to reach their designated final acceleration of 10 m/s².  The thrust required to maintain this constant acceleration is given initially by:

$$T_B = (M_{ship} + m_{thread})a \text{ and } T_C = M_{ship}a. \tag{1}$$

The designations B and C refer respectively to spaceships B and C.  The difference arises because ship B is tugging the thread while the thread initially has no effect on the trailing ship C.

The tension on the right-hand side of the thread is given by:

$$T_{thread} = m_{thread}a = (0.05 kg)\left(10\ \frac{m}{s^2}\right) = 0.5\ N. \tag{2}$$

This is substantially less than the 10 N force required to break the fragile thread.  If the thread is too fragile to survive this initial acceleration, it is too fragile to test Bell's paradox.

The distance between the ships remains constant at one kilometer in the rest frame of ship A.  But the proper length of the thread in the frame moving with respect to ship A is continuously increasing by the factor $1/\sqrt{1 - v^2/c^2}$.  The thread breaks when it has stretched by 7% to 1070 m.  The velocity corresponding to this amount of stretch is found by solving Eq. (3):

$$\text{Proper Length of Thread} = \frac{1000\ m}{\sqrt{1 - \frac{v^2}{c^2}}} = 1070\ m. \quad (3)$$

The thread breaks when the ships are moving in the rest frame of ship A at v = 0.26c = 7.7 x 10⁷ m/s.

Not so Fast

There is a practical difficulty with the scenario described above. As the thread tries to contract in the rest frame, it applies a tiny retarding tension to the leading spaceship and a small boost to the trailing ship. In order for the two ships to continue accelerating at <u>exactly</u> 10 m/s², the thrusts, $T_B$ and $T_C$, have to continuously be adjusted, $T_B$ up and $T_C$ down. These adjustments in thrust are very, very small but absolutely necessary to keep the two ships accelerating in lockstep.

Real engines cannot instantaneously adjust their thrusts to keep the acceleration of a spaceship equal to some predetermined value. Assume the engines in this experiment are designed to keep the acceleration of each ship within 0.1 % of 10 m/s². Then the acceleration of each ship will stay between 9.999 and 10.001 m/s². It turns out that this is not nearly good enough! It will be shown later that the acceleration will need to be kept in the range $(10 \pm 10^{-12})$ m/s² for Bell's Paradox to proceed without being disrupted.

Imagine the thread replaced by a rod to make the image more substantial. Accelerating a classical rod by pushing or pulling it does not change the length of the rod because the push or pull is transmitted through the rod with an infinite velocity. Every part of the rod instantaneously feels the push or pull. Signals in special relativity cannot travel at infinite speeds. In special relativity pushing a rod to accelerate it becomes problematic because the other end of the rod does not instantaneously know that it is being pushed. The pushed end of the rod starts moving before the other end. That time difference introduces tensions in the rod.

To avoid those time differences, both ends of the rod need to begin accelerating at the same instant. Bell's paradox demonstrates that when both ends accelerate at exactly the same rate, tensions still develop because the rod or thread wants to contract. The rod has served its purpose and the discussion returns to the thread connecting the two ships.

For the thread to accelerate without any internal tensions, ship B, needs to accelerate more slowly than ship C. Then at any given instant in the initial rest frame, ship C is moving faster than ship B. The correct difference in accelerations makes the slope of the line connecting the two ships when both are moving with velocity v equal to v/c². This line is by construction one of constant time, Δt' = 0, in the instantaneously co-moving inertial frame of spaceships B and C.

The relationship between the proper acceleration of the front ship, $a_B$, and that of the back ship, $a_C$, that satisfies the above condition is given by[3]:

$$\frac{1}{a_B} - \frac{1}{a_C} = \frac{L}{c^2}. \tag{4}$$

Re-write Eq. (4) in terms of the differential acceleration δa:

$$\frac{1}{a} - \frac{1}{a + \delta a} = \frac{L}{c^2}. \tag{5}$$

Solve this equation for δa:

$$\delta a = a \left( \frac{\frac{aL}{c^2}}{1 - \frac{aL}{c^2}} \right) \cong \frac{a^2 L}{c^2} \tag{6}$$

For this *gedanken* experiment the value of $aL/c^2$ is approximately $10^{-13}$ leading to a value of $10^{-12}$ m/s² for the differential acceleration required to keep the proper length of the thread 1 kilometer. This is a very, very small difference in the accelerations of the two spaceships: much smaller than the assumed feedback system that was designed to keep the acceleration of each ship between 9.999 and 10.001 m/s².

The amount of tension in the thread required to change the acceleration of one of the ships by $10^{-12}$ m/s² is:

$$\delta T \cong M_{ship}\, \delta a = (5 \times 10^6 kg)\left(10^{-12}\frac{m}{s^2}\right) = 5 \times 10^{-6} N. \tag{7}$$

In the spirit of generating ballpark figures, assume the force needed to stretch the thread is linear over the range from $5 \times 10^{-6}$ N to 10 N. That assumption leads to the following ratio for the amount of stretch, x, caused by a tension of $5 \times 10^{-6}$ N:

$$\frac{x}{5 \times 10^{-6} N} = \frac{70\ m}{10\ N} \rightarrow x = 0.035\ mm. \tag{8}$$

The velocity of the two ships increases until δT becomes just large enough to produce the differential acceleration needed to move the thread as a rigid body. That event corresponds to the proper length of the thread increasing by 0.035 mm. The velocity necessary to produce that increase is given by:

$$1000\ m = (1000 + 0.000035\ m)\sqrt{1 - \frac{v^2}{c^2}}. \tag{9}$$

The solution of Eq. (9) is v = 80,000 m/s. When the ships reach this velocity, tiny tensions in the thread begin to produce differential accelerations on the order of $10^{-12}$ m/s$^2$. This is just enough to decrease the acceleration of the spaceship in front and increase the acceleration of the ship in back by the amount needed to accelerate the thread as a rigid body. Once the thread has stretched 0.035 mm, $\delta T$ causes the ship C to accelerate faster that ship B in just enough to keep the proper length of the string 1 km plus 0.0035 mm. The string will never break independent of then speed of the two ships with respect to the rest frame defined by ship A.

Revisit Bell's Paradox

The following is a qualitative description of the motion of the two spaceships connected by a fragile thread attempting to keep their respective proper accelerations equal to 10 m/s$^2$:

*The two ships accelerate without incident until they reach speeds of about 80,000 m/s. At this point, the tension in the string is large enough to create a differential acceleration that is approximately equal to difference that allows the string to maintain a constant proper length. That differential acceleration is $\delta a_{ideal}$.*

*Because of random fluctuations in the thrust of the engines, $\delta a$ changes from moment to moment. As long as $\delta a = \delta a_{ideal}$, the ships continue to move without affecting the tension in the thread. If $\delta a < \delta a_{ideal}$, the thread begins to stretch increasing the tension in the thread. The added tension speeds up ship C and impedes ship B to return $\delta a$ to its optimal value. On the other hand if $\delta a > \delta a_{ideal}$, the tension in the thread decreases causing $\delta a$ to decrease and causing $\delta a$ drift back to its optimal value.*

These self-correcting actions keep the two spaceships moving in a way that maintains the tension in the thread on the order of $10^{-12}$ N, ten trillion times less than the tension needed to break the fragile thread! In this *gedanken* experiment testing Bell's paradox, the thread never breaks and the ships can accelerate forever[4].

Conclusion

This hypothetical test of Bell's Spaceship Paradox is based on the <u>assumption</u> that it is not possible to have engines that can continuously adjust their thrusts to keep two spaceships moving with an acceleration between $(10 \pm 10^{-12})$ m/s$^2$. Without this assumption, Bell's Paradox is true as stated.

---

[1] Edward M. Dewan and Michael J. Beran, "*Note on stress effects due to relativistic contraction,*" Am. J. Phys **27**, 517-518, (1959).

[2] J.S. Bell, *Speakable and Unspeakable in Quantum Mechanics*, Cambridge University Press (1987).

[3] Jerold Franklin, *Foundations of Physics*, **43**, Issue 12, 1489-1501, (2013)

[4] If the thread is stout enough to survive the initial acceleration, it is extremely unlikely that it will break because of any relativistic contractions suffered as the ships continue their accelerating motion.